\documentclass{article}
\usepackage{spconf,amsmath,graphicx}
\usepackage{stfloats}
\usepackage{bm}
\usepackage{amssymb}
\usepackage{mathtools}
\usepackage{tabularx}
\usepackage{multirow}
\usepackage{algorithm}
\usepackage{algpseudocode}
\usepackage{changepage}
\usepackage{enumitem}

\title{semi-supervised multi-channel speaker diarization with cross-channel attention}
\name{
\begin{tabular}{c}
Shilong Wu\textsuperscript{1}, Jun Du\textsuperscript{1,*}\thanks{\textsuperscript{*}corresponding author}, Maokui He\textsuperscript{1}, Shutong Niu\textsuperscript{1}, Hang Chen\textsuperscript{1}, \\ Haitao Tang\textsuperscript{2}, Chin-Hui Lee\textsuperscript{3}
\end{tabular}
} 
\address{
 \textsuperscript{1} University of Science and Technology of China, China \\ \textsuperscript{2} iFlytek, China  
   \textsuperscript{3} Georgia Institute of Technology, USA
}
\begin{document}
\ninept
\maketitle
\begin{abstract}
Most neural speaker diarization systems rely on sufficient manual training data labels, which are hard to collect under real-world scenarios. This paper proposes a semi-supervised speaker diarization system to utilize large-scale multi-channel training data by generating pseudo-labels for unlabeled data. Furthermore, we introduce cross-channel attention into the Neural Speaker Diarization Using Memory-Aware Multi-Speaker Embedding (NSD-MA-MSE) to learn channel contextual information of speaker embeddings better. Experimental results on the CHiME-7 Mixer6 dataset which only contains partial speakers' labels of the training set, show that our system achieved 57.01\% relative DER reduction compared to the clustering-based model on the development set. We further conducted experiments on the CHiME-6 dataset to simulate the scenario of missing partial training set labels. When using 80\% and 50\% labeled training data, our system performs comparably to the results obtained using 100\% labeled data for training.

\end{abstract}
\section{Introduction}
\label{sec:intro}

Speaker diarization refers to the process of labeling speech timestamps with classes corresponding to speaker identity \cite{dia_review}. As a crucial component in speech processing systems, speaker diarization has garnered substantial research attention and finds extensive application in various domains, including conference transcription, telephone conversation analysis, and other scenarios \cite{dia_review2}. However, due to the complex acoustic scenes, a large number of speech overlaps, and the lack of sufficient labeled data, speaker diarization still faces great challenges \cite{M2Met}.

Traditional speaker diarization systems are based on clustering methods, mainly involving speaker feature extraction and clustering. The first step involves extracting the speaker's representation \cite{i-vector,x-vector} from the input speech segment, while the second step utilizes various clustering algorithms such as spectral clustering (SC) \cite{sc}, k-means clustering \cite{kmeans}, mean shift clustering \cite{mean} and agglomerative hierarchical clustering (AHC) \cite{AHC,AHC2}, to aggregate the regions of each speaker into separate clusters. Although some systems have achieved good performance \cite{clu, VBx}, clustering methods are constrained by their capability only to assign a single speaker label to each speech segment, making it difficult to handle overlapping speech segments. In recent years, researchers have paid more attention to end-to-end neural speaker diarization (EEND) systems \cite{e2e2019,e2e20192,e2etsvad,e2e2022}, which treat the diarization task as a multi-label classification problem, enabling the model to handle overlapping speech effectively. Recently, target-speaker voice activity detection (TS-VAD) \cite{ts-vad} was proposed. It utilizes speech features and speaker embedding as input to directly predict each speaker’s activeness for each frame, showing good performance. Based on this, Neural Speaker Diarization using Memory-Aware Multi-Speaker Embedding (NSD-MA-MSE) \cite{MA-MSE} has been proposed as an innovative speaker diarization network. It introduces a dedicated memory module and utilizes the attention mechanism to extract cleaner and more discriminative multi-speaker embeddings from memory, showing superior performance. 

Nevertheless, the efficacy of most neural speaker diarization systems heavily depends on the availability of sufficient training data. The challenges arise from the huge cost of annotating data in real-world scenarios and the limited performance exhibited by models trained on alternative datasets. As a semi-supervised learning method, the pseudo-label \cite{pseudo1} strategy is a good choice to handle this problem, which can generate estimated labels for unlabeled data to utilize larger-scale training data. In speech recognition, this method has been applied to improve performance \cite{semiasr,semiasr2,semiasr3}. For speaker diarization, \cite{semi2} proposes a system to generate pseudo-labels for unlabeled data, but it requires a seed model trained on additional datasets, which can lead to increased computational costs and the potential domain mismatch issue when significant differences exist among the datasets. In this paper, we propose a novel system: semi-supervised multi-channel speaker diarization with cross-channel attention. We assume that part of the data is labeled and the other part is unlabeled in the training set. This assumption is common and pragmatic for real-world scenarios. For example, only the interviewee's speech is labeled in the interview, or in other scenarios (e.g., meetings), only part of the sessions are labeled to reduce annotation costs.


Firstly, we propose a strategy to generate pseudo-labels for unlabeled data. We first train the time-delay neural network based speech activity detection (TDNN-SAD) \cite{TDNN-SAD} model using near-field audio of labeled data in the training set and decode the near-field audio of unlabeled data utilizing the trained model to obtain speaker-wise initial pseudo-labels. Then we use labeled and unlabeled data (with initial pseudo-labels) to train a neural speaker diarization model and use the trained model to decode far-field audio to obtain the refined pseudo-labels. In addition, we also propose an effective multi-channel neural speaker diarization model. We introduce cross-channel attention to learn contextual relationships across channels to better process the multi-channel audio data. Our inspiration is derived from \cite{multiasr}, with the distinction that our attention module is fed with the embeddings of each speaker rather than the features of each speech frame. We introduced this module into the NSD-MA-MSE \cite{MA-MSE} system, resulting in Multi-Channel Neural Speaker Diarization Using Memory-Aware Multi-Speaker Embedding (MC-NSD-MA-MSE). This extension enables the system to extract and fuse speaker embeddings from diverse channels while preserving the inherent advantages of the original system. In our study, the speaker diarization model utilizes our proposed MC-NSD-MA-MSE model in both the pseudo-label generation and testing phases. We conducted experiments on Mixer6 and CHiME-6 datasets of the CHiME-7 DASR Challenge \cite{chime7}. The performance comparison of the system and the impact of different proportions of labeled data on the results are discussed respectively. Experimental results show the effectiveness of our system. Finally, we also explore the impact of iteration times on the results and the results of the speaker diarization system on ASR.

\section{Proposed method: semi-supervised multi-channel speaker diarization}
\label{sec:method}

In this section, we will provide a detailed explanation of the method to generate pseudo-labels for unlabeled data in subsection 2.1, and the Multi-Channel NSD-MA-MSE model in subsection 2.2.

\vspace{-0.1cm}
\subsection{Method to generate pseudo-label}
\label{ssec:semi}
Based on the presence or absence of labels, the training set is divided into two categories: labeled data and unlabeled data. The labels refer to the timestamps of each speaker's speech segments. We aim to generate pseudo-labels for unlabeled data, obtaining complete labels for the training dataset. Afterward, we can use the complete labeled training set to train the speaker diarization model.

Figure 1 illustrates the main procedure of the method to generate pseudo-labels for unlabeled data. In addition to far-field audio, our method assumes the availability of near-field audio, which is common in data recording.

The system can be seen as a two-stage system and each stage involves training and decoding steps. The whole process can be summarized as the following 4 steps:
\vspace{-10pt}
\begin{adjustwidth}{1.62em}{}
\item[1)]
TDNN-SAD model training: Train the time-delay neural network-based speech activity detection (TDNN-SAD) model using near-field audio of the labeled data.
\end{adjustwidth}
\vspace{-13pt}
\begin{adjustwidth}{1.62em}{}
\item[2)]
Initial pseudo-labels generation: Perform speech activity detection to the near-field audio for each speaker of the unlabeled data utilizing the TDNN-SAD model trained in step 1. The timestamps of detected speech segments serve as the preliminary annotation for the corresponding speaker's speech segments, functioning as the initial pseudo-labels for the unlabeled data.
\end{adjustwidth}
\vspace{-13pt}
\begin{adjustwidth}{1.62em}{}
\item[3)]
End-end neural speaker diarization model training: Train an end-end neural speaker diarization model using far-field audio of labeled and unlabeled data (with initial pseudo-labels). The choice of speaker diarization model is flexible. Here, we use our proposed MC-NSD-MA-MSE model.
\end{adjustwidth}
\vspace{-13pt}
\begin{adjustwidth}{1.62em}{}
\item[4)]
Final pseudo-labels generation: Utilize the MC-NSD-MA-MSE model trained in step 3 to decode the far-field data of unlabeled data in the training set, obtaining a set of diarization results. Treat the obtained results as the more accurate pseudo-labels for unlabeled data.
\end{adjustwidth}

In the following subsections, two key parts of our system will be elaborated in detail.

\vspace{-0.1cm}
\subsubsection{Using TDNN-SAD model to generate initial pseudo-labels}
\label{sssec: Initialization labels generation}
To obtain initial pseudo-labels for unlabeled data, we use the speech activity detection (SAD) model with near-field audio. Since near-field audio predominantly captures the speech activity of the current speaker, the timestamps of the speech segments detected by the SAD model can serve as the diarization label for that particular speaker. The SAD model we used is based on time-delay neural networks, namely TDNN-SAD. TDNN model captures longer context information through time \cite{TDNN}, which can effectively improve the performance of SAD and has a certain degree of noise robustness. The structure of TDNN-SAD is similar to \cite{vad}. A major difference in our system is that we perform local normalization on the extracted features. This normalization process proves beneficial in mitigating energy variations and minimizing the influence of noise across speech segments, thereby improving the accuracy of identifying speech activity boundaries. We first extract Mel-frequency Cepstral Coefficient (MFCC) features for the input audio: $\bm{\mathrm{X}} = [\bm{\mathrm{x}}_1,\bm{\mathrm{x}}_2,...,\bm{\mathrm{x}}_T]$, where $\bm{\mathrm{x}_t}\in\mathbb{R}^{D}$ is the $D$-dimensional $(D=40)$ MFCC feature of the $t$-th frame and $T$ is the frame number of the current audio. 

\begin{figure}[t]
\centering
\includegraphics[width=\linewidth]{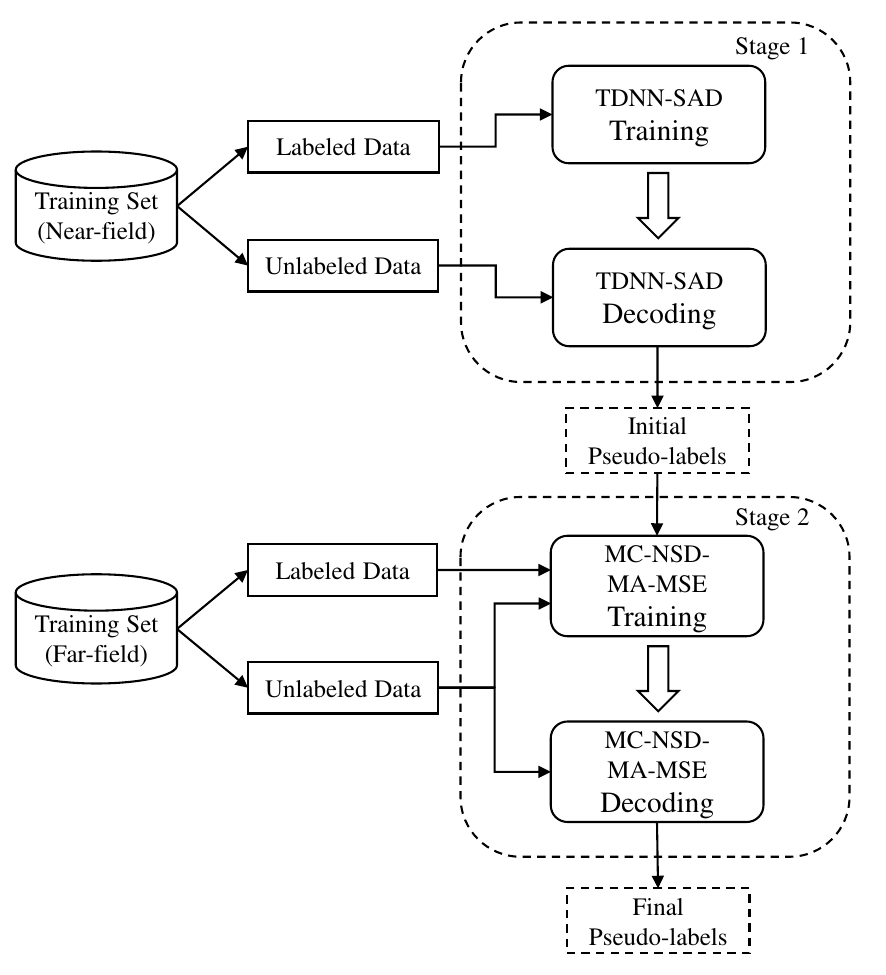}
\vspace{-0.7cm}
\caption{Flow of our proposed method to generate pseudo-labels for unlabeled data}
\label{semi}
\vspace{-0.4cm}
\end{figure}

These features are first normalized and subsequently fed into a sequence of 5 layers of TDNN, followed by 2 layers of statistics pooling \cite{TDNN-2}. 
The network is designed with an overall context duration of approximately 1 second, consisting of approximately 0.8 seconds of left context and 0.2 seconds of right context. The output can be represented as  ${\hat{Y}} = [\hat{y_1},\hat{y_2},...,\hat{y_T}]$, where $\hat{y_t}$ is the probability of speech at $t$-th frame. The network is trained using the cross-entropy loss function to predict the speech/non-speech labels.
\vspace{-3pt}
\begin{equation}
L(y_t,\hat{y_t}) = -[{y_t}\rm{log}(\hat{y_\mathit{t}})+(1-{y_\mathit{t}})\rm{log}(1-\hat{y_\mathit{t}})]
\end{equation}
where $y_t$ are the label of the $t$-th frame.

In the post-processing stage, we utilize the hidden Markov models (HMMs) for Viterbi decoding to find the most likely sequence of speech activities. This method effectively exploits the temporal dependencies between consecutive frames and captures the holistic structure of speech activity. Finally, we can obtain the result $s_t$ to determine whether a frame is speech.
\begin{equation}
s_t=\begin{cases} 1& \text{the $t$-th frame is speech}\\0& \text{the $t$-th frame is not speech} \end{cases}
\end{equation}

In step 1, the TDNN-SAD model is trained using near-field audio of the labeled data. In step 2, the trained model is used as a pre-trained model to decode the speaker-wise near-field audio of the unlabeled data and determine the presence of speech activity at each frame. Given that near-field audio primarily consists of the speech of a single speaker, aggregating the SAD results from each speaker's near-field audio in a session provides a rough estimation of the diarization labels for the unlabeled data, namely initial pseudo-labels.

\vspace{-0.1cm}
\subsubsection{Using MC-NSD-MA-MSE model to generate final pseudo-labels}
\label{sssec: Estimated labels generation}
In the process of recording near-field speech, inevitably, a few instances of other speakers' voices may also be captured, resulting in many false alarm (FA) errors in SAD results. Therefore, relying solely on the SAD to generate the pseudo-labels may not guarantee complete reliability (we will explain this in the analysis of experimental results). To obtain more accurate pseudo-labels, we will proceed to steps 3 and 4. By incorporating the initial pseudo-labels of the unlabeled data obtained from the previous step, we can obtain all the labels of the entire training set. We use multi-channel far-field data to train a neural speaker diarization model. The use of far-field data is to take advantage of multi-channel information and to avoid the lack of diarization results 
caused by incomplete near-field audio collection due to factors like movement or walking. The choice of speaker diarization model is flexible. In this paper, we use a novel speaker diarization model with cross-channel attention mechanism. This model is based on the Neural Speaker Diarization using Memory-Aware Multi-Speaker Embedding (NSD-MA-MSE) \cite{MA-MSE}, which further uses the multi-channel information to learn the contextual relationships of speaker embeddings across channels, called Multi-Channel NSD-MA-MSE (MC-NSD-MA-MSE). The specific structure of this model will be elaborated in section 2.2. 

The MC-NSD-MA-MSE model trained in step 3, is utilized to decode the multi-channel far-field audio of the unlabeled data in step 4. Then, speaker diarization results are obtained, serving as the final pseudo-labels for the unlabeled data. At this stage, we have completed labeling the training set. To make the pseudo-labels more accurate, we can iterate through steps 3 and 4 to obtain new pseudo-labels. We will discuss the effectiveness of doing so in the experimental section.

\begin{figure*}[htbp]
\centering
\includegraphics[width=1\textwidth]{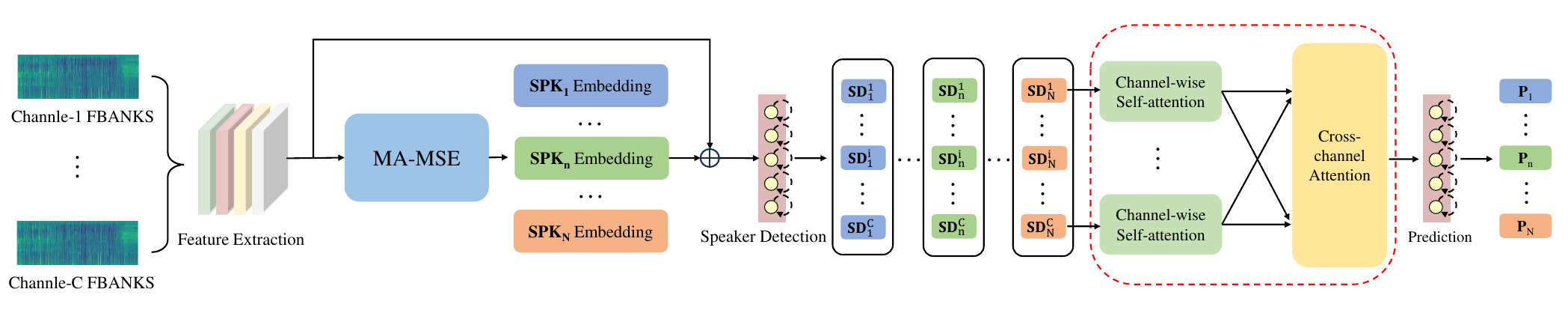}
\vspace{-0.6cm}
\caption{The proposed Multi-Channel NSD-MA-MSE model, assuming that there are $C$ channels of far-field audio and $N$ speakers}
\label{multi-channel}
\vspace{-0.4cm}
\end{figure*}

\subsection{Multi-Channel NSD-MA-MSE model}
\label{ssec:multi-channel}

Different speaker positions and angles are included in multi-channel audio, providing more rich and comprehensive signals. This helps to improve the accuracy and robustness of speech analysis and processing, and can reduce the impact of environmental noise, echoes, and other interference. Driven by this motivation, based on NSD-MA-MSE, we introduced the cross-channel attention mechanism to utilize the multi-channel audio data.

NSD-MA-MSE can extract clearer and more distinctive multi-speaker embeddings from memory through the attention mechanism, unlike TS-VAD \cite{ts-vad}, which simply extracts i-vectors from roughly estimated speaker segments. This method can extract more accurate information for each speaker on overlapping segments and make the quality of speaker embeddings more robust, greatly improving the performance of the speaker diarization. On this basis, our proposed MC-NSD-MA-MSE retains the advantages of the original system while effectively utilizing the information present in multi-channel audio data. The architecture is shown in Figure 2, consisting of three parts, namely, the main network, the memory-aware multi-speaker embedding (MA-MSE) module, and the cross-channel attention module. The signals of each channel need to be separately processed through the main network and the MA-MSE module, then fused in the cross-channel attention module, and finally output the results of the speaker diarization. In the following sections, we will elaborate on the structure of each part.

\subsubsection{Main network}
\label{sssec: Main network}

The main network consists of a frame-level feature extraction module consisting of four convolutional layers, a speaker detection module consisting of two-layer bidirectional long short-term memory with projection (BLSTMP), and a prediction module consisting of one-layer BLSTMP. The input of the system is a set of FBANKs, and the frame-level depth features extracted by CNNs are represented as the matrix $\bm{\mathcal{F}} = [\bm{\mathrm{F}}_1,...,\bm{\mathrm{F}}_i,...,\bm{\mathrm{F}}_C]$, where $\bm{\mathrm{F}_i}\in\mathbb{R}^{T\times D}$ is the $D$-dimensional frame-level feature vector of the $i$-th channel, $T$ is the frame number of the utterance, and $C$ is the channel number of the audio. The frame-level features are inputs for the main network and the MA-MSE module. Furthermore, the frame-level features are concatenated with a collection of speaker embeddings $\bm{\mathrm{E}} = [\bm{\mathrm{SPK}}_1,...,\bm{\mathrm{SPK}}_i,...,\bm{\mathrm{SPK}}_C]$ from the MA-MSE module by copy for all $T$ frames, where $\bm{\mathrm{SPK}_i}\in\mathbb{R}^{N\times L}$ is the $L$-dimensional vector of the speaker for $i$-th channel, $N$ is the number of the speakers. Then, deeper frame-level speaker-wise features are extracted from each speaker concatenation using the Speaker Detection model. The features are fed into a multi-channel attention module, integrated with intra-channel and cross-channel contextual information to derive a new set of features. Subsequently, the $N$ speaker features are concatenated and supplied to the prediction module, which produces $N$ two-class outputs $\bm{\hat{Y}}=[\bm{\mathrm{P}}_1,...,\bm{\mathrm{P}}_n,...,\bm{\mathrm{P}}_N]$, where $\bm{\mathrm{P}}_n = [\hat y_n(1),..., \hat y_n(t),..., \hat y_n(T)]$ represents the probabilities of speech and silence of speaker $n$ in each frame.

\subsubsection{MA-MSE module}
\label{sssec: MA-MSE}
The memory-aware multi-speaker embedding (MA-MSE) module can extract the more discriminative multi-speaker embeddings from the memory block through the attention mechanism. The key component of MA-MSE is the memory block, where memory refers to a set of speaker embedding bases inspired by the dictionary learning concept \cite{dictionary}. It contains the i-vectors or x-vectors extracted from additional datasets, and the speaker embedding bases vectors in memory are easily distinguished.

The inputs to the MA-MSE module are the frame-level features $\bm{\mathcal{F}}$ from the main network and the speaker mask matrix $\bm{\mathcal{M}} = [\bm{\mathrm{M}}_1,...,\bm{\mathrm{M}}_i,...,\bm{\mathrm{M}}_C]$, where $\bm{\mathrm{M}_i}\in\mathbb{R}^{N\times T}$ denotes whether each speaker is speaking in each frame of the $i$-th channel audio. We use a clustering-based model to decode the audio to obtain $\bm{\mathcal{M}}$. By multiplying $\bm{\mathcal{M}}$ by $\bm{\mathcal{F}}$, we can obtain the frame-level features of each speaker as input for the attention module. Based on these features and the speaker embedding bases, the attention mechanism is used to select the speaker embedding base that is most similar to the current speech segment from each memory and then combine them into a vector called aggregated speaker vector as the output $\bm{\mathrm{E}}$.

\subsubsection{Cross-channel attention module}
\label{sssec: attention}
To better utilize multi-channel information, we introduce a cross-attention module, which includes two main blocks, a channel-wise self-attention block and a cross-channel attention block. This method has been applied in ASR \cite{multiasr}, but previous applications have been based on audio features \cite{multiasr2}. In contrast, our approach applies the method at the level of speaker-wise features. Through this method, we can pay more attention to the speaker-wise feature and combine the contextual information within the channel and the information between channels to obtain a more representative and robust representation vector. To our best knowledge, this is the first time this method has been applied to the speaker diarization system. 

The frame-level features of the $n$-th speaker extracted after the speaker detection module on the $i$-th channel are represented as $\bm{\mathrm{SD}}_n^i$. We apply channel-wise self-attention to the deep embeddings $\bm{\mathrm{SD}}_n^i$ of each speaker in each channel, outputting a more accurate speaker expression that combines contextual information within the single channel. Then, we perform cross-channel attention on the specific speaker across all channels to obtain a set of final representations for speakers: $\bm{\mathcal{S}} = [\bm{\mathrm{S}}_1,...,\bm{\mathrm{S}}_n,...,\bm{\mathrm{S}}_N]$.

\textbf{Channel-wise self-attention:}
We use multi-head attention to obtain the weights across time within the single channel. The queries, keys, and values are obtained from the $n$-th speaker embeddings $\bm{\mathrm{SD}}_n^i$ of the $i$-th channel by the following function:

\vspace{-0.1cm}
\begin{equation}
\begin{split}
\bm{\mathrm{Q}}_{i,n}^{\mathrm{S}} = \sigma (\bm{\mathrm{SD}}_i^n\bm{\mathrm{ W}}^{\mathrm{S},q}+\bm1(\bm{\mathrm{b}}_{i}^{\mathrm{S},q})^T)\\
\bm{\mathrm{K}}_{i,n}^{\mathrm{S}} = \sigma (\bm{\mathrm{SD}}_i^n\bm{\mathrm{ W}}^{\mathrm{S},k}+\bm1(\bm{\mathrm{b}}_{i}^{\mathrm{S},k})^T)\\
\bm{\mathrm{V}}_{i,n}^{\mathrm{S}} = \sigma (\bm{\mathrm{SD}}_i^n\bm{\mathrm{ W}}^{\mathrm{S},v}+\bm1(\bm{\mathrm{b}}_{i}^{\mathrm{S},v})^T)
\end{split}
\end{equation}
where $\sigma$ is the ReLU activation function, $\bm{\mathrm{W}},\bm{\mathrm{b}}$ are the learnable weights and bias parameters, and S represents using attention-wise self-attention to each single channel. The output is computed by:

\begin{equation}
\bm{\mathrm{H}}_{i,n}^{\mathrm{S}} = \mathrm{Softmax}\left(\frac{\bm{\mathrm{Q}}_{i,n}^{\mathrm{S}}(\bm{\mathrm{K}}_{i,n}^{\mathrm{S}})^T}{\sqrt{d_m}}\right)\bm{\mathrm{V}}_{i,n}^{\mathrm{S}}
\end{equation}
where the scaling $\sqrt{d_m}$ is for numerical stability. We then obtain the final output $\hat{\bm{\mathrm{H}}_{i,n}^{\mathrm{S}}}$ through the feed-forward layers.

\textbf{Cross-channel attention:}
Based on the self-attention output of each speaker in each channel, we use the cross-channel attention module to learn the contextual relationships of the speaker features across channels and time steps. The calculation of queries is similar to the previous step, but Keys and Values utilize information from other channels, i.e., the average contribution of other channels:

\vspace{-0.1cm}
\begin{equation}
\begin{split}
\begin{aligned}
&\bm{\mathrm{Q}}_{i,n}^{\mathrm{M}} = \sigma (\hat{\bm{\mathrm{H}}_{i,n}^{\mathrm{S}}}\bm{\mathrm{W}}^{\mathrm{M},q}+\bm1(\bm{\mathrm{b}}_{i}^{\mathrm{M},q})^T)\\
&\bm{\mathrm{K}}_{i,n}^{\mathrm{M}} = \sigma (\bm{\mathrm{H}}_{i,n}^{\mathrm{S'}}\bm{\mathrm{W}}^{\mathrm{M},k}+\bm1(\bm{\mathrm{b}}_{i}^{\mathrm{M},k})^T)\\
&\bm{\mathrm{V}}_{i,n}^{\mathrm{M}} = \sigma (\bm{\mathrm{H}}_{i,n}^{\mathrm{S'}}\bm{\mathrm{W}}^{\mathrm{M},v}+\bm1(\bm{\mathrm{b}}_{i}^{\mathrm{M},v})^T)
\end{aligned}
\end{split}
\end{equation}

\begin{equation}
\bm{\mathrm{H}}_{i,n}^{\mathrm{S'}} = \frac{1}{C} \sum_{j,j\neq i}\hat{\bm{\mathrm{H}}_{j,n}^{\mathrm{S}}}
\end{equation}
where C is the number of channels, and M represents using cross-channel attention to multi-channel. The cross-channel attention output is then computed by:
\begin{equation}
\bm{\mathrm{H}}_{i,n}^{\mathrm{M}} = \mathrm{Softmax}\left(\frac{\bm{\mathrm{Q}}_{i,n}^{\mathrm{M}}(\bm{\mathrm{K}}_{i,n}^{\mathrm{M}})^T}{\sqrt{d_m}}\right)\bm{\mathrm{V}}_{i,n}^{\mathrm{M}}
\end{equation}
Then we obtain the final output $\hat{\bm{\mathrm{H}}_{i,n}^{\mathrm{M}}}$ through the feed-forward layers. Finally, we integrate the outputs generated by all channels and use a global average pooling layer to obtain the final feature representation of each speaker $\bm{\mathcal{S}}=[\bm{\mathrm{S}}_1,...,\bm{\mathrm{S}}_n,...,\bm{\mathrm{S}}_N]$.
\begin{equation}
\bm{\mathrm{S}}_n=\frac{1}{C}\sum_{i=1}^C\hat{\bm{\mathrm{H}}_{i,n}^{\mathrm{M}}}
\end{equation}

By connecting the above results and inputting them into the final prediction module of the main network, we can obtain the speaker diarization results.

\section{Experimental setup}
\label{sec: Experiments and results}

\subsection{Evaluation data}
\label{ssec:Dataset}
We evaluate the proposed method using two datasets, namely, the Mixer 6 Speech dataset and the CHiME-6 dataset. The division of the training, development (DEV) and evaluation (EVAL) sets follows the standards of the CHiME-7 DASR Challenge \cite{chime7}.

The Mixer 6 Speech dataset comprises 594 unique native English speakers engaged in a collection of 1425 sessions, each with an approximate duration of 45 minutes. The dataset includes a total of 13 channels of audio data, of which channels 1, 2, and 3 are dedicated to capturing near-field audio using close-talk microphones. Our study uses the $train\_intv$ set as the training set, which is partitioned as the CHiME-7 DASR Challenge. Significantly, in that two-speaker interview conversation setup of the training set, only the timestamps for the subject's speech are labeled, while the timestamps for the interviewer's speech remain unlabeled. Using the proposed system, we estimate the pseudo-labels for the interviewer's speech timestamps and evaluated the performance in the development (DEV) and evaluation (EVAL) set.

The CHiME-6 dataset focuses on real-home environments, where each session involves 4 participants. The dataset is characterized by challenging acoustic conditions and exhibits a high rate of speech overlap, with 23\% in the training set and 43.8\% in the development set, bringing enormous difficulties for the speaker diarization task. The dataset includes near-field speech recordings acquired through individual speakers wearing binaural microphones and far-field speech captured by 6 Kinect array devices. Each array is equipped with 4 microphones. In our study, to simulate the scenario of missing partial training set labels, we randomly select 20\%, 50\%, and 80\% of the training set data as the labeled data, while the remaining training set data is considered unlabeled, and we evaluate the performance in the development set. This approach allowed us to investigate the influence of different proportions of the labeled training set data on system performance in real-world scenarios.

\vspace{-0.15cm}
\subsection{Baseline system}
\label{ssec:Baseline}

\vspace{-0.05cm}
\subsubsection{Clustering-based and TS-VAD based system}
In order to better show the performance of our method, we use the clustering-based speaker diarization model as the baseline. We first use the ECAPA-TDNN model \cite{ECAPA} to extract the x-vector for each speech segment. Then we choose the spectral clustering algorithm to cluster speaker embeddings and get the diarization results. In addition, the TS-VAD model is also used in the first experiment as the baseline. we follow the method in \cite{tsvad}, which is proposed to handle the unknown number of speakers. 

\vspace{-0.2cm}
\subsubsection{Single channel NSD-MA-MSE based system}

To explore the performance of our proposed multi-channel model, we compared it with the original single-channel-based NSD-MA-MSE system, called SC-NSD-MA-MSE. We follow a similar system configuration in \cite{MA-MSE} and use 128 classes i-vectors as the speaker embedding in the memory block. In order to make the comparison fairer, we also used the DOVER-Lap \cite{dover-lap} method for the single-channel NSD-MA-MSE system, which fuses the results from all channels to use the channel information. This method is also used in the clustering-based and TS-VAD based systems mentioned above.

\vspace{-0.1cm}
\subsection{Model configuration}
\label{ssec: model configuration}
For the Multi-Channel NSD-MA-MSE system, the input is the 40-dim FBANKs feature. In the cross-channel attention module, we used 8 attention heads. The configuration of the MA-MSE block is consistent with SC-NSD-MA-MSE. We also used the post-processing strategy in \cite{ustc} to merge speech frames with short pauses. In the training stage, we use Adam optimizer with a learning rate of 0.0001 to optimize the model on 8 NVIDIA Tesla A100 GPUs and apply the model of the $4$-th epoch. In the testing stage, the results of the clustering-based model are used to get the speaker mask matrix.

In order to evaluate the impact of the results generated by the speaker diarization system on ASR, we added a speech recognition model as the backend. We first use guided source separation (GSS) \cite{GSS} to separate speech using the results of diarization. For ASR, We adopted an encoder-decoder structure based on the attention mechanism. Among them, the encoder adopts a 12-layer transformer, and the decoder includes an embedding layer, a 6-layer transformer, and an output layer. In order to better extract features, we adopt a more robust self-supervised pre-training model, namely, wavLM, which is trained based on 94k hours of unlabeled data. We also add a speech enhancement (SE) model based on the Conv-TasNet \cite{SE}. Finally, fine-tune the entire network (SE+ASR) \cite{ASR}.

\subsection{Evaluation metric}
\label{ssec: evaluation metrics}
We use the diarization error rate (DER) \cite{DER} to evaluate the accuracy of the systems. The lower DER is the better. For our DER computation, we evaluated all of the errors, including overlapping speech segments. And the DER is calculated with a 250 ms collar, following the setup in CHiME-7 DASR Challenge.

To evaluate the influence of speaker diarization on speech recognition, we employ Diarization Attributed WER (DA-WER) \cite{chime7} as the metric for ASR. Specifically, the Hungarian method is used to obtain the optimal mapping that minimizes the DER between the reference and hypothesis segments. We use the official evaluation script provided by CHiME-7 DASR Challenge to calculate DA-WER.

\begin{table}[t]
\vspace{-0.2cm}
\caption{DERs (\%) and DA-WERs (\%) comparison of different models on DEV set of Mixer 6 with pseudo-labels of different stages.}
\vspace{5pt}
\setlength\tabcolsep{8.23pt}
\renewcommand{\arraystretch}{1.2}
\begin{tabular}{cccc}
\hline
Model                        & \multicolumn{1}{l}{Pseudo-label} & DER   & DA-WER \\ \hline
Clustering                           & -               & 16.19 & 19.43  \\ \hline
\multirow{2}{*}{TS-VAD} & Initial                         & 13.97 & 16.62  \\
                              & Final                         & 12.79 & 15.80  \\ \hline
\multirow{2}{*}{SC-NSD-MA-MSE} & Initial                         & 11.20 & 15.69  \\
                              & Final                         & 10.42 & 15.12  \\ \hline
\multirow{2}{*}{MC-NSD-MA-MSE} & Initial                          & 8.21 & 13.05  \\
                              & Final                        & \textbf{6.96} & \textbf{12.44}  \\ \hline
\end{tabular}
\vspace{-0.5cm}
\end{table}

\vspace{-0.1cm}
\section{RESULTS AND ANALYSIS}
\label{sec: RESULTS AND ANALYSIS}

We evaluate our semi-supervised system for two different situations. The evaluation on Mixer 6 is for the case where only some speakers in each session of the training set are labeled, while other speakers in this session are unlabeled. The evaluation on CHiME-6 is for the case where some sessions in the training set are fully labeled and others are completely unlabeled. We first use our system to generate pseudo-labels for unlabeled data, then use labeled and pseudo-labeled data to retrain the MC-NSD-MA-MSE model, and evaluate the DER jointly results on the DEV or EVAL set. We also explored the impact of iteration times on the results. In addition, we also use the ASR model to measure DA-WER to observe the impact of speaker diarization results on speech recognition.

\vspace{-0.2cm}
\subsection{Evaluation on Mixer 6 }
\label{ssec: mixer6}

In the training set of Mixer 6, each session has two speakers, and only one speaker's speech is labeled. Therefore, it is necessary to generate pseudo-labels for another unlabeled speaker by our semi-supervised system. We can then use the labeled and the unlabeled data (with pseudo-labels) to train the model and evaluate the results. Table 1. shows the DER comparison of different diarization methods on the Mixer 6 DEV set. The clustering-based model directly utilizes far-field data for clustering, while TS-VAD, SC-NSD-MA-MSE and MC-NSD-MA-MSE are not only used for pseudo-label generation in stage 2, but also for final model training and evaluation. In addition, we also explore the impact of pseudo-labels generated in different stages on the results. 'Initial' represents training the model using the initial pseudo-labels generated by TDNN-SAD, and 'Final' represents training the model using the final pseudo-labels.

\begin{figure}[t]
\centering
\includegraphics[width=\linewidth]{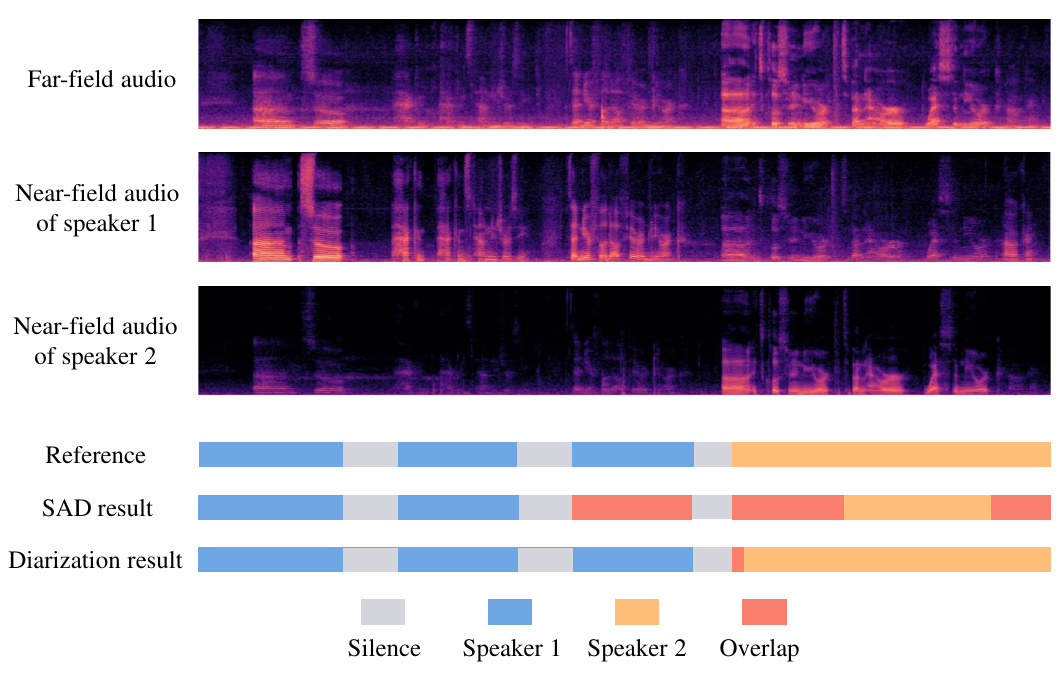}
\vspace{-0.8cm}
\caption{An example of comparing SAD result with speaker diarization result. SAD result refers to decoding the speaker-wise near-field audio by SAD and combining the results. Diarization result refers to decoding far-field audio using MC-NSD-MA-MSE model}
\label{example}
\vspace{-0.1cm}
\end{figure}

The results show that our proposed MC-NSD-MA-MSE based system exhibits the best performance, achieving a DER of 6.96\% on the DEV set, which is a 57.01\% relative DER reduction compared to the clustering-based model. Compared with SC-NSD-MA-MSE, which employs DOVER-Lap to fuse all channels on the result, the relative reduction of DER is 33.21\%, demonstrating the superiority of using cross-attention for speaker embeddings. Additionally, training the model using the final pseudo-labels yields better results than the initial pseudo-labels, validating the necessity of employing the speaker diarization module to refine the pseudo-labels. The reason is that when the near-field audio contains obvious voices from other speakers, the SAD result may contain many false alarms, as shown in Figure 3, resulting in inaccurate initial false labels. But using the MC-NSD-MAMSE model with far-field audio can effectively improve this problem.

We also explored the effect of the number of iterations on the results of the Mixer 6 EVAL set as shown in Table 2. The iteration here refers to the speaker diarization model trained with pseudo-labels to decode the far-field audio of the training set again, generate new pseudo-labels and retrain the model. The results show that the best results (DER of 5.70\%) of EVAL set can be achieved in the second iteration. This means that our system does not need to spend too much time to achieve good results.

\begin{table}[t]
\vspace{-0.3cm}
\caption{Detailed DERs (\%) of the results on EVAL set of Mixer 6 with different iterations to generate pseudo-labels. DER is composed of False alarms (FA), Misses (MISS), and Speaker errors (SPKERR).}
\vspace{5pt}
\setlength\tabcolsep{6.6pt}
\renewcommand{\arraystretch}{1.2}
\begin{tabular}{cccccc}
\hline
Model                                                                   & Iterations & FA   & MISS  & SPKERR & DER   \\ \hline
Clustering                                                                      &   -     & 0.11 & 14.19 & 3.46   & 17.76 \\ \hline
\multirow{3}{*}{\begin{tabular}[c]{@{}c@{}}MC-NSD-\\ MA-MSE\end{tabular}} & 1      & 2.56 & 3.25  & 0.21   & 6.02 \\
                                                                         & 2      & 2.39 & 3.13  & 0.18   & \textbf{5.70} \\
                                                                         & 3      & 2.42 & 3.17  & 0.19   & 5.78 \\ \hline
\end{tabular}
\vspace{-0.4cm}
\end{table}

\subsection{Evaluation on CHiME-6}
\label{ssec: chime6}
In order to explore the impact of the amount of unlabeled data on the results in real-world scenarios, we conducted experiments using the CHiME-6 dataset to simulate the situation where part of the data in the training set is unlabeled. As shown in Table 3, we randomly selected 80\%, 50\%, and 20\% of the sessions in the CHiME-6 training set as labeled data, while the remaining sessions are regarded as unlabeled. For each case, we conducted two experiments: one involved training the MC-NSD-MA-MSE model directly using the labeled data of that percentage, while the other involved generating pseudo-labels for the unlabeled data using our semi-supervised system and training the model using both labeled and pseudo-labeled data. The results are reported on the CHiME-6 development set.

The results show that using 100\% of the training data to train the MC-NSD-MA-MSE model can achieve a DER of 30.76\% on DEV set. However, when only using a portion of the training data to directly train the model (as shown as 'Direct' in the table), significant performance degradation is observed. Therefore, the amount of training data substantially impacts the model's performance. When using our semi-supervised system to generate pseudo-labels for unlabeled data before training (as shown as 'Semi-supervised' in the table), we found that in the case of 50\% and 80\% of the data with labels, the results obtained can match (DER of 30.77\%) or surpass (DER of 30.42\%) the performance achieved using 100\% labeled data. The reason for the improvement is that manual labeling may lead to errors, and our method can correct some of these errors. 
However, due to the presence of a large number of other speakers' voices in near-field audio of the CHiME-6 dataset, using SAD to generate initial pseudo-labels may result in some incorrect labels, which leads to performance degradation when using 20\% of labeled training data. If the near-field audio is cleaner, the system will perform better. Nevertheless, it is also greatly improved compared to direct training. The above results show that our method is feasible on partially labeled real data.

\begin{table}[t]
\vspace{-0.25cm}
\caption{Comparison of DERs (\%) and DA-WERs (\%) for different proportions of labeled training set data and different training methods (direct or semi-supervised) on DEV set of CHiME-6.}
\vspace{5pt}
\setlength\tabcolsep{6.5pt}
\renewcommand{\arraystretch}{1.2}
\begin{tabular}{ccccc}
\hline
Model                        & Data         & Training method & DER   & DA-WER \\ \hline
Clustering                           &   -                   &  -               & 38.77 & 43.49  \\ \hline
\multirow{7}{*}{\begin{tabular}[c]{c@{}}MC-\\ NSD-\\ MA-MSE\end{tabular}} & 100\%                 & Direct          & 30.76 & 35.67  \\ \cline{2-5} 
                              & \multirow{2}{*}{80\%} & Direct          & 32.34 & 36.38  \\
                              &                      & Semi-supervised & 30.42 & 35.15  \\ \cline{2-5} 
                              & \multirow{2}{*}{50\%} & Direct          & 34.32 & 39.51  \\
                              &                      & Semi-supervised & 30.77 & 35.63  \\ \cline{2-5} 
                              & \multirow{2}{*}{20\%} & Direct          & 38.28 & 41.97  \\
                              &                      & Semi-supervised & 32.24 & 36.46  \\ \hline
\end{tabular}
\vspace{-0.4cm}
\end{table}

\subsection{Impact of speaker diarization results on ASR}
\label{ssec: ASR}
The speaker diarization system generates speech timestamps for each speaker, which the ASR model utilizes to transcribe speaker-attributed speech. The accuracy of the speaker diarization system directly impacts the alignment between transcriptions and individual speakers, consequently improving the overall transcription accuracy. Table 1 shows a strong correlation between DA-WER and DER, achieving the lowest DA-WER of 12.44\% at the lowest DER. Table 3 also shows that our method can also achieve comparable results with 100\% labeled training data. Therefore, by using our semi-supervised system to improve the speaker diarization results, we can effectively improve the accuracy of the speaker-attributed speech recognition as well.

\vspace{-0.1cm}
\section{conclusion}
\label{sec: cnoclusion}

In this paper, we propose a novel semi-supervised speaker diarization system to solve the problem of missing labels in real-world scenarios by generating pseudo-labels on unlabeled data. In addition, we also introduce the cross-channel attention mechanism to effectively utilize the multi-channel information. The efficacy of our method has been demonstrated through experiments conducted on the Mixer 6 and CHiME-6 datasets in CHiME-7 DASR Challenge. We also analyzed the impact of speaker diarization results on ASR performance. In the future, we will expand the semi-supervised system to the unsupervised system for wider applications.

\vfill
\pagebreak

\bibliographystyle{IEEEbib}
\bibliography{refs}

\end{document}